\documentclass[draft]{agujournal}

 \usepackage{colortbl}
 \usepackage{multirow}
  \usepackage{graphicx}
  \usepackage{color}
  \usepackage{amssymb}
  \usepackage{caption}
  \usepackage{rotating}

\journalname{Geophysical Research Letters}

\begin{document}

%
%


\title{Assessing the time dependence of reconnection with Poynting's theorem: MMS observations}

%
%




\authors{K.~J. Genestreti\affil1,
	      P.~A. Cassak\affil2,
	      A. Varsani\affil3,
	      J.~L. Burch\affil4,
	      R. Nakamura\affil1,
	      S. Wang\affil5}

 \affiliation{1}{Space Research Institute, Austrian Academy of Sciences, Graz, Austria}
 \affiliation{2}{West Virginia University, Morgantown, West Virginia, USA}
 \affiliation{3}{Mullard Space Science Laboratory, University College London, Dorking, UK}
 \affiliation{4}{Southwest Research Institute, San Antonio, Texas, USA}
 \affiliation{5}{University of Maryland, College Park, MD, USA}





\correspondingauthor{Kevin J. Genestreti}{kevin.genestreti@oeaw.ac.at}




\begin{keypoints}
\item Each term in Poynting's theorem is calculated for an MMS observed electron diffusion region (EDR)
\item The equality of both sides of the equation shows that the terms are accurately determined
\item Magnetic energy accumulation is observed at the electron current sheet around the X-point
\end{keypoints}

%
%


\begin{abstract}
We investigate the time dependence of electromagnetic-field-to-plasma energy conversion in the electron diffusion region of asymmetric magnetic reconnection. To do so, we consider the terms in Poynting's theorem. In a steady state there is a perfect balance between the divergence of the electromagnetic energy flux $\nabla \cdot \vec{S}$ and the conversion between electromagnetic field and particle energy $\vec{J} \cdot \vec{E}$. This energy balance is demonstrated with a particle-in-cell simulation of reconnection. We also evaluate each of the terms in Poynting's theorem during an observation of a magnetopause reconnection region by Magnetospheric Multiscale (MMS). {{We take the equivalence of both sides of Poynting's theorem as an indication that the errors associated with the approximation of each term with MMS data are small.}} We find that, for this event, balance between $\vec{J}\cdot\vec{E}=-\nabla\cdot\vec{S}$ is only achieved for a small fraction of the energy conversion region at/near the X-point. Magnetic energy was rapidly accumulating on either side of the current sheet at roughly three times the predicted energy conversion rate. {{ Furthermore, we find that while $\vec{J}\cdot\vec{E}>0$ and $\nabla\cdot\vec{S}<0$ are observed, as is expected for reconnection, the energy accumulation is driven by the overcompensation for $\vec{J}\cdot\vec{E}$ by $-\nabla\cdot\vec{S}>\vec{J}\cdot\vec{E}$.}} We note that due to the assumptions necessary to do this calculation, the accurate evaluation of $\nabla\cdot\vec{S}$ may not be possible for every MMS-observed reconnection event; but if possible, this is a simple approach to determine if reconnection is or is not in a steady-state.
\end{abstract}

%
%

%


%
%
%
%

\section{Introduction}

Observations of magnetopause reconnection by Magnetospheric Multiscale (MMS) have revealed that energy conversion can occur in highly localized regions of the electron diffusion region (EDR) much more rapidly than previously expected \cite{Burch.2016b,BurchandPhan.2016,Ergun.2016b,Ergun.2017,Chen.2017,Hwang.2017,Genestreti.2017}. This energy conversion rate is often expressed {{as the work rate of the non-ideal electric field, or}} $\vec{J}\cdot(\vec{E}+\vec{v}_e\times\vec{B})\equiv\vec{J}\cdot\vec{E}'$ \cite{Zenitani.2011}, where $\vec{J}$ is the current density, $\vec{E}$ is the electric field, $\vec{B}$ is the magnetic field, and $\vec{v}_e$ is the electron bulk velocity. \cite{Cassak.2017} analyzed 2.5-d particle-in-cell (PIC) simulations of three of these EDR events. They found that the MMS-observed energy conversion rates were up to several orders of magnitude larger than what was seen in their simulation for laminar steady-state reconnection. One explanation offered by \citeauthor{Cassak.2017} is that the large energy conversion rates {{observed by}} MMS may constitute localized bursts of activity in time and/or space rather than the global rate. Given the ubiquity with which larger-than-predicted $\vec{J}\cdot\vec{E}'$ are observed in magnetopause EDRs, it is possible that spatial and/or temporal burstiness is also ubiquitous.

To further investigate the steadiness or burstiness of energy conversion in the EDR, we consider Poynting's theorem, which in differential form is

\begin{equation}
-\frac{\partial u}{\partial t}=\nabla\cdot\vec{S}+\vec{J}\cdot\vec{E},
\end{equation}

\noindent where $u=(\epsilon_0E^2+B^2/\mu_0)/2$ is the electromagnetic energy density and $\vec{S}=\vec{E}\times\vec{B}/\mu_0$ is the Poynting vector. {{(Note that in Poynting's theorem, the energy conversion rate $\vec{J}\cdot\vec{E}$ contains both ideal $\vec{J}\cdot(-\vec{v}_e\times\vec{B})$ and non-ideal $\vec{J}\cdot\vec{E}'$ terms.)}} The rate of change of the electromagnetic energy density $\partial u/\partial t$ is zero in any perfectly {{steady-state process, in which case the power exerted by the electric field on the particles $\vec{J}\cdot\vec{E}$ is balanced by the net electromagnetic energy flux into a volume $\nabla\cdot\vec{S}$ at every point in space}}. When integrated over a volume, Poynting's theorem says the net electromagnetic energy flux into a volume is balanced by the power exerted by the particles on the fields in the entire volume. For example, in an idealized (2-d, laminar, steady-state, and symmetric) reconnection ion diffusion region (IDR), the net difference between the electromagnetic energy densities flowing into the reconnection site $E_MB_L/\mu_0$ and expelled from the reconnection site $E_MB_N/\mu_0$ balances the rate of energy conversion within the reconnection site {{contributed mainly by}} $J_ME_M$. In this coordinate system, $\pm$L is the direction of the reconnecting magnetic fields, N is the current sheet normal in the reconnection plane, and M completes the right-handed coordinate system. \textit{Swisdak et al.} [submitted] suggested the dissipation of $B_M$ may also occur within the asymmetric EDR, given the presence of a non-zero $J_NE_N$. With MMS, it is possible to {{approximate}} each of the terms in Poynting's theorem to assess the extent to which reconnection is steady-state.

In this letter we present the first experimental determination of equation (1) in the context of reconnection (to our knowledge). To put our results in context, we also analyze the terms in Poynting's theorem in a 2.5-d PIC simulation of reconnection. Simulation results are discussed in Section 2. In Section 3 we describe our methodology for evaluating equation (1) from the MMS data, likely sources of errors, and results. Section 4 concludes with a discussion.

\section{Poynting's theorem from a 2.5-d PIC simulation} 

We analyze the results of a fully kinetic PIC simulation that was
carried out and analyzed in \cite{Cassak.2017}.  The
simulations and their setup are thoroughly discussed in that
reference, so we only provide the most salient details here.  The code
in use is the P3D code \cite{Zeiler.2002}.  The initial setup of
the simulation was chosen to match the upstream conditions of an
asymmetric reconnection EDR event with an order 1 guide field observed
by MMS on 8 December 2015 \cite{BurchandPhan.2016,Genestreti.2017}, 
which is not the same event as what is studied in Section 3
but has a comparable out-of-plane (guide) magnetic field strength.
The simulation was shown to reproduce some key features observed by
MMS, e.g., partially-formed crescent-shaped electron distribution
functions \cite{Hesse.2014,Hesse.2016,Burch.2016b} and a
non-ideal energy conversion rate that was peaked between the X and
electron stagnation points.

The simulation was in two dimensions with doubly periodic boundary
conditions in a rectangular domain of size 40.96 $\times$ 20.48 with a
grid scale of 0.01 in units of $d_{i0}$, the ion inertial length with
respect to a density of $n_0 = 15$ cm$^{-3}$.  The reference magnetic
field strength was 35 nT.  The upstream values on the magnetosheath
side were $B_L = 0.429, B_M = 0.4, n = 0.5, T_i = 1.313$, and $T_e =
0.123$; upstream values on the magnetospheric side were $B_L = 1, B_M
= 0.357, n = 0.2, T_i = 1.361$, and $T_e = 0.271$.  {{The full set of upstream 
parameters are listed in physical units and normalized values in Table 1 of \cite{Cassak.2017},
under the column header ``8 Dec 2015''.}} The electron mass
is 100 times smaller than the ion mass and $c / c_{A0} = 25$.  The
initial setup uses a double tanh profile for the magnetic fields and
temperatures, with a density profile chosen to impose MHD force
balance.  There are initially an average of 500 particles per grid
with equal weight.  A small coherent perturbation is used to initiate
reconnection.  Lengths and times are presented in terms of reference
values on the magnetosheath side, the inertial scale $d_{i,sh}$ and
the inverse ion cyclotron frequency $\Omega_{ci,sh}^{-1}$,
respectively.

\begin{figure*}
\centering
\noindent\includegraphics[width=13pc]{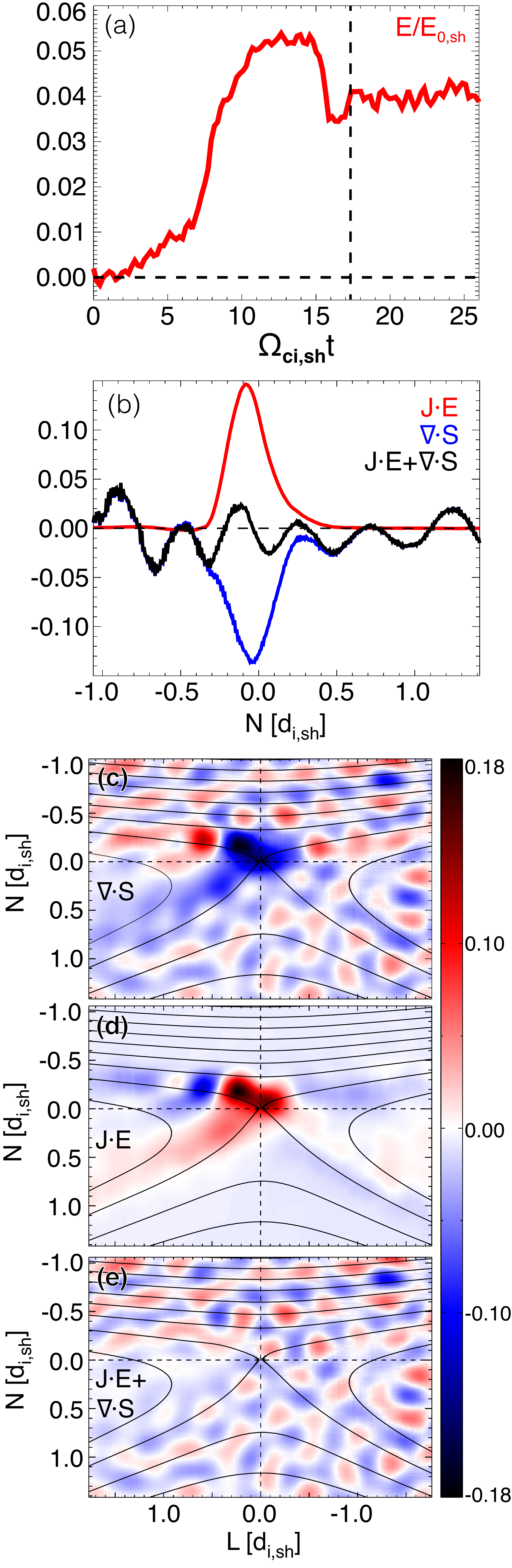}
\caption{(a): The reconnection rate normalized to an electric field of 1.8 mV/m as a function of time normalized to $\Omega_{ci,sh}$, the inverse ion cyclotron frequency evaluated upstream in the magnetosheath. The vertical line shows the time where we calculate the quantities in (b)--(e). (b): The terms in Poynting's theorem evaluated along a normal-directed cut through the X-point. (c), (d), and (e): $\nabla\cdot\vec{S}$, $\vec{J}\cdot\vec{E}$, and $\nabla\cdot\vec{S}+\vec{J}\cdot\vec{E}$ (respectively) over the portion of the 2-d simulation domain surrounding the EDR. Length scales are normalized by $d_{i,sh}$, the asymptotic ion inertial length in the upstream magnetosheath. The solid black lines in (c)--(e) are contours of the magnetic flux function, i.e., magnetic field lines.}
\label{sim}
\end{figure*}

We consider a time in the simulation, $\Omega_{ci,sh} t = 17.16$ as shown in the vertical line in Figure \ref{sim}a,  where reconnection has evolved and is progressing at a more or less constant reconnection rate $E$, as shown as a function of time $t$. $\vec{J}\cdot\vec{E}$, $\nabla\cdot\vec{S}$, and $\vec{J}\cdot\vec{E}+\nabla\cdot\vec{S}$ are shown in Figure \ref{sim}b along a normal-directed cut through the X-point and also over a 2-d domain in Figures \ref{sim}c, d, and e, respectively. To reduce the noise in the Poynting vector divergence term that always arises when taking derivatives of noisy data, we average the simulation results in time over $\sim0.09$ $\Omega_{ci,sh}$ (10 time steps) and smooth the results in the spatial domain by $\sim$2 $d_{e,sh}$ (or $\sim$0.2 $d_{i,sh}$, which is 20 grid cells). Despite these efforts, fluctuations in $\nabla\cdot\vec{S}$ are observed with wavelengths near the smoothing length. However, these fluctuations are generally less than $\sim$30\% as large as the largest value of $\nabla\cdot\vec{S}$ at and very near the X-point, as seen in Figure \ref{sim}b and c.

There is a strong $\vec{J}\cdot\vec{E}>0$ in the center of the reconnection region (see Figures \ref{sim}b and d). This data does not show noisy fluctuations because no derivatives needed to be taken to obtain $\vec{J}\cdot\vec{E}$. The strong positive $\vec{J}\cdot\vec{E}$ is co-located with a strong influx of electromagnetic energy, i.e., $\nabla\cdot\vec{S}<0$ (Figures \ref{sim}b and c). These two terms balance one another to within the noise level of $\nabla\cdot\vec{S}$ (Figures \ref{sim}b and e). This energy balance appears to be achieved beyond the center of the reconnection region as well, as $\vec{J}\cdot\vec{E}>0$ and $\nabla\cdot\vec{S}<0$ appear together in the magnetosheath-side ($N>0$) separatrix of the northern ($L>0$) reconnection region, {{and $\vec{J}\cdot\vec{E}<0$ and $\nabla\cdot\vec{S}>0$ appear together in the magnetosphere-side separatrix of the northern exhaust}}. This result is not unexpected for 2-d steady-state laminar reconnection.

\section{Poynting's theorem from MMS observations}

On 28 November 2016, MMS crossed very slowly from the magnetosphere to the magnetosheath through an EDR [\textit{Genestreti et al.}, submitted]. The separation between the MMS probes was close to the smallest separation used by MMS to date, at 6.4 km or $\sim4.5$ $d_{e,sh}$. When MMS crossed the magnetosphere-side separatrix, it was in the southern outer EDR or the IDR. MMS then moved northward into the central EDR, where non-ideal energy conversion ($\vec{J}\cdot\vec{E}'>0$) was observed near the X-point. Here, all four spacecraft were simultaneously within the $\vec{J}\cdot\vec{E}'>0$ region. Likely as a result of the slow crossing and small inter-probe separation, \textit{Genestreti et al.} [submitted] found that the divergence and curl terms in the generalized Ohm's and Ampere's laws were very well resolved by the four-probe linear gradient technique \cite{ISSIchap14}. Furthermore, they found evidence that the structure of the current sheet and electric field may have been consistent with the 3-d and turbulent picture of asymmetric reconnection \cite{Price.2016,Price.2017}, rather than the 2-d and laminar picture. They based this conclusion on observations of large in and out-of-the-reconnection plane electron pressure forces in the central EDR, as net pressure forces out-of-the reconnection plane cannot occur in the 2-d picture.

We use the highest time resolution data from MMS, including the DC magnetic field vector measured by the fluxgate magnetometers \cite{Russell.2016}, the coupled AC-DC electric field vector from the electric field double probes \cite{Ergun.2016,Lindqvist.2016}, and the plasma electron moments from the fast plasma investigation \cite{Pollock.2016}. We have to make a number of assumptions in order to approximate the terms in Poynting's theorem. (1) We assume the fields and plasma moments vary linearly within the volume of the tetrahedron. (2) We smooth the AC-DC electric field data to obtain a DC field. We assume that $\nabla\cdot\vec{S}_{AC}$ and $\partial u_{AC}/\partial t$ (which cannot be resolved with the linear gradient technique) do not affect the balance of energy on the scale of the MMS tetrahedron. 

\begin{figure*}
\centering
\noindent\includegraphics[width=39pc]{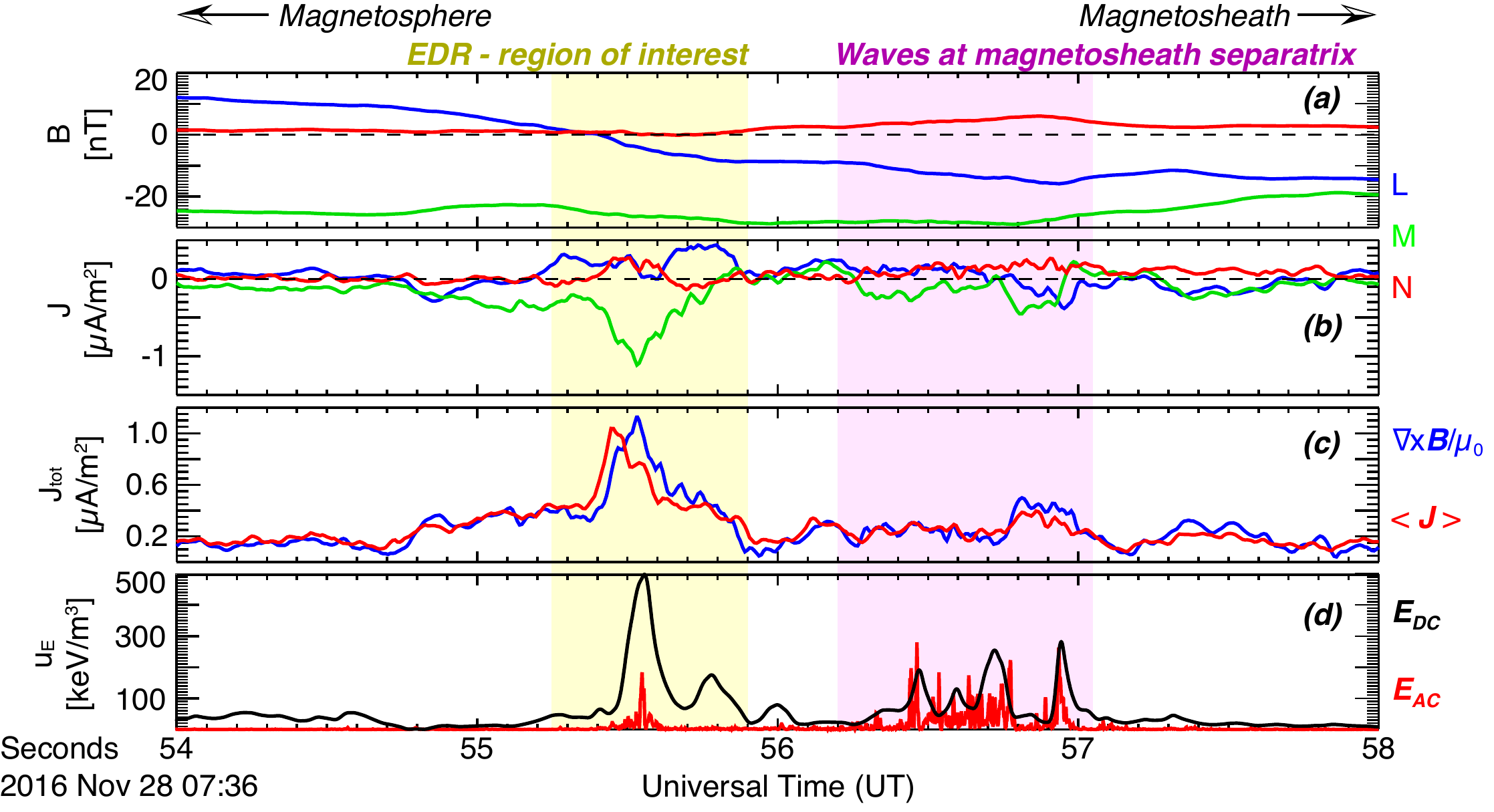}
\caption{(a): Magnetic field vector and (b): current density vector in the LMN coordinate system defined for the X-point in \textit{Genestreti et al.} [submitted]. (c): Total current density from the curlometer (blue) and averaged plasma moments (red). (d): The energy stored in the DC (low-frequency, black) and AC (high-frequency, red) portions of the electric field.}
\label{qual}
\end{figure*}

First, we check the validity of these assumptions. Figure \ref{qual} shows MMS data near the intense out-of-plane electron current layer around the X-point. The region of intense current, non-ideal energy conversion, electron agyrotropy, anisotropic electron heating, etc., is highlighted in yellow. The magnetosheath separatrix is in pink. As is shown in Fig. \ref{qual}c, the barycentric current density vector can be calculated nearly identically using either the curlometer technique (blue) {{or}} the 4-spacecraft-averaged plasma moments data (red). This is an indication that the variation of the DC magnetic field may have been approximately linear within the tetrahedron volume. \textit{Genestreti et al.} [submitted] was also able to accurately calculate the electron pressure divergence term in generalized Ohm's law, which is an indication that the variations in the plasma electron moments were approximately linear. Figure \ref{qual}d shows that the energy density of the AC electric field is small compared to the energy density in the DC field in the EDR. Here, the DC field is determined by smoothing the coupled AC-DC electric field over 0.1 seconds and the AC field is defined as the difference of the two. Given the speed of the magnetopause was estimated to be 31 km/s [\textit{Genestreti et al.}, submitted] and $d_{e,sh}\approx1.4$ km, this 0.1-s smoothing time corresponds to a distance of roughly 2 $d_{e,sh}$, which is comparable to the spatial smoothing used to analyze our PIC simulation. As is shown in Figure \ref{qual}d, the energy in the AC field only becomes comparable to the energy of the DC field in the separatrix (pink boxed) region. For this reason, we focus our investigation solely on the EDR.

We do not expect that the terms in equation (1) can be obtained for every MMS EDR event. In some cases, dissipation appears to be driven nearly entirely by high-frequency and intensely localized electric waves \cite{Burch.2017} [\textit{Swisdak et al.}, submitted], {{such that the MMS tetrahedron may not resolve gradients over the associated scale sizes}}. In other cases, the spacecraft separation is larger than the size of the structure and the linear gradient technique cannot clearly and closely reproduce features of the reconnection region, e.g., the dissipation rate \cite{Torbert.2016b,Genestreti.2017}. As such, we define a ``good quality'' estimation of the terms in equation (1) as one in which both sides of the equation, which are determined separately, are roughly equivalent. I.e., if $\partial u/\partial t$ and the errors associated with its calculation are roughly equal to $-\vec{J}\cdot\vec{E}-\nabla\cdot\vec{S}$ and their associated errors then we assume that the error terms are small.

\begin{figure*}
\centering
\noindent\includegraphics[width=32pc]{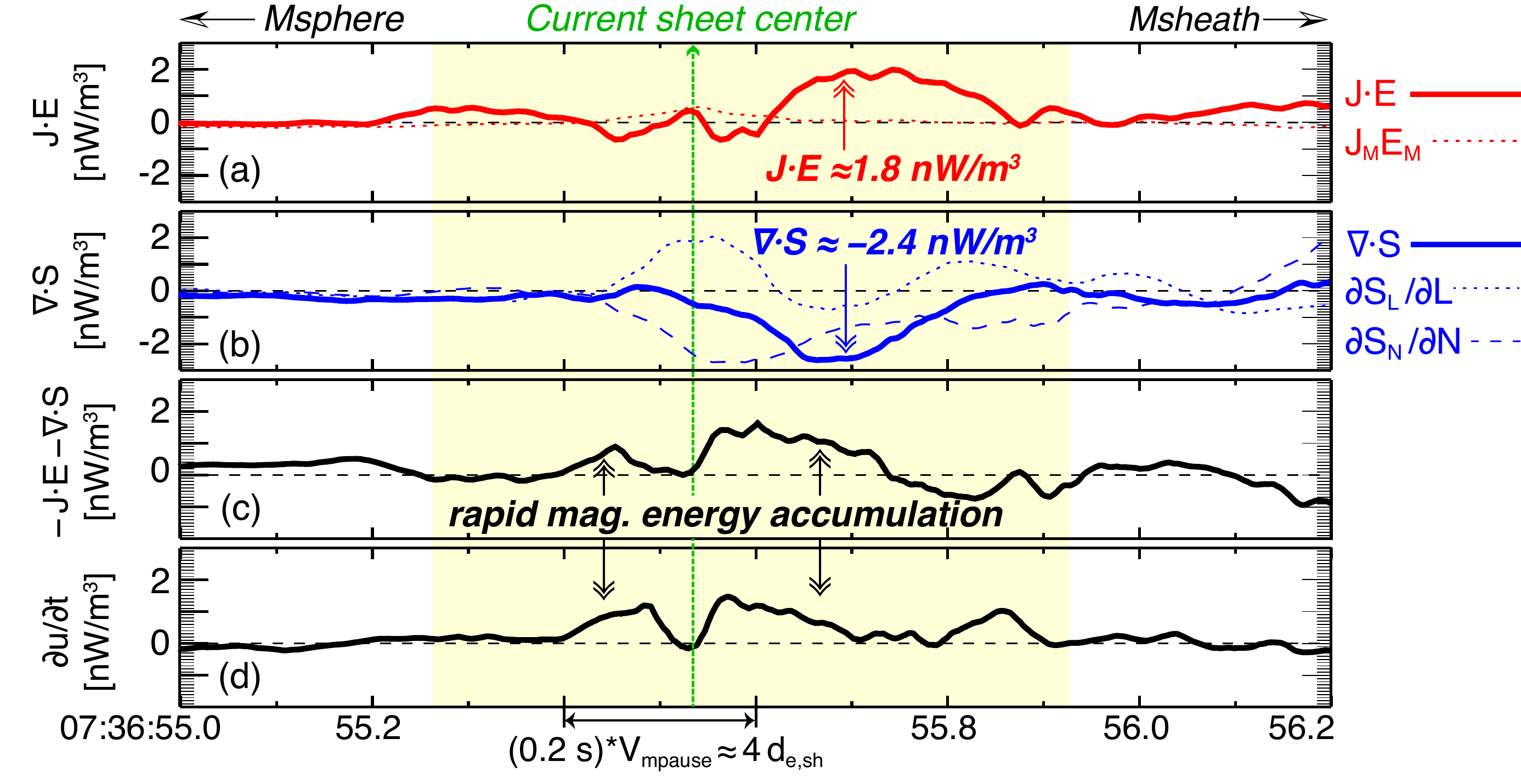}
\caption{(a): The energy conversion rate $\vec{J}$$\cdot$$\vec{E}$ (solid line) and the rate of work of the reconnection electric field $J_ME_M$ (dotted). (b): The energy flux divergence $\nabla$$\cdot$$\vec{S}$ (solid) and the energy inflow rate $\partial S_N/\partial N$ (dashed) and outflow rate $\partial S_L/\partial L$ (dotted). (c) and (d): The per-volume rate of change of the electromagnetic energy density, as determined by the right and left-hand sides of Poynting's theorem (Equation 1), respectively.}
\label{pt}
\end{figure*}

The terms in equation (1) are shown for the 28 November event in Figure \ref{pt}. $\vec{J}\cdot\vec{E}$, shown in Fig. \ref{pt}a, is determined by the inner product of the curlometer current \cite{ISSIchap14} and the 4-point-averaged smoothed (DC) electric field in the spacecraft frame. (There are no significant differences between the spacecraft frame and the magnetopause frame).  $\nabla\cdot\vec{S}$, shown in Fig \ref{pt}b, is determined by taking the linear divergence of the cross product of the DC electric and magnetic fields. According to our quality criterion, $-\vec{J}\cdot\vec{E}-\nabla\cdot\vec{S}$ (Fig \ref{pt}c) should be equivalent to our independently calculated $\partial u/\partial t$ (Fig \ref{pt}d). Here, $\partial u/\partial t$ is calculated as $\partial u/\partial t=du/dt-\vec{v}_e\cdot\nabla u$, where the full time derivative is determined from the time series of 4-point-averaged DC fields and the electron velocity $\vec{v}_e$ is used in the convective derivative. For all terms, $\nabla$ is approximated with the linear gradient technique \cite{ISSIchap14}.

First we note that the left (Fig. \ref{pt}d) and right-hand (Fig. \ref{pt}c) sides of equation (1) match extraordinarily well, which was our ultimate quality criterion. The largest value of $\vec{J}\cdot\vec{E}\approx1.8$ nW/m$^3$ is dominated by the action of the in-plane electric field components $E_L$ and $E_N$ on the plasma. The largest energy conversion and influx rates are also observed at 7:36:55.7 UT between the reconnection mid-plane (7:36:55.5 UT) and the magnetosheath-side separatrix (7:36:56.4 UT), rather than at the center of the current sheet. Also, while $\vec{J}\cdot\vec{E}$ and $\nabla\cdot\vec{S}$ balance one another in the center of the current sheet, they are not balanced elsewhere. Both methods for calculating $\partial u/\partial t$ show that the electromagnetic energy density is increasing on either side of the current sheet. While the electric field intensity may also be changing in time, this energy density increase appears to be almost entirely from an increase in the magnetic energy density. The largest value of $\partial u/\partial t\approx1.5$ nW/m$^3$ is comparable in magnitude to the largest value of $\vec{J}\cdot\vec{E}\approx1.8$ nW/m$^3$ and is nearly three times larger than the predicted value of $J_ME_{M,pred}\approx0.6$. Overall, we conclude that the EDR was not in a steady state at the time when and place where it was observed by MMS. {{Furthermore, we note that the cause of the energy accumulation is an overcompensation for the field-to-plasma energy conversion by more rapid energy influx (i.e., $-\nabla\cdot\vec{S}>\vec{J}\cdot\vec{E}>0$).}}

Note that balance between $\vec{J}\cdot\vec{E}$ (Fig \ref{pt}a) and $\nabla\cdot\vec{S}$ (Fig \ref{pt}b) is achieved at the current sheet center near/at the X-line, which is marked by the vertical dashed green line in Fig \ref{pt}. The energy conversion due to the reconnection electric field $J_ME_M$ (dashed red line in Fig \ref{pt}a) is largest at the center of the current sheet. The value of $J_ME_M\approx0.7$ nW/m$^3$ is also almost identical to the predicted maximum value for steady-state reconnection with a reconnection rate of 0.1. Given the upstream conditions for this event listed in Table 1 of \textit{Genestreti et al.} [submitted] and the Cassak-Shay formula for the asymmetric reconnection rate \cite{CassakandShay.2007}, the predicted reconnection electric field is $E_{M,pred}\approx0.5$ mV/m. Together with the observed maximum current density of $J_M\approx1.1$ $\mu$A/m$^2$, this yields a predicted energy conversion rate of $J_ME_{M,pred}\approx-0.6$ nW/m$^3$. Also as expected, more electromagnetic energy enters the central current than is expelled from it, as $\partial S_N/\partial N\approx-2.4$ nW/m$^3$, $\partial S_L/\partial L\approx1.9$ nW/m$^3$, and because of a small but finite $\partial S_M/\partial M$, we find $\nabla\cdot\vec{S}\approx0.6$ nW/m$^3$. All of this leads to balance between the energy conversion and Poynting flux divergence terms at the X-point, where $-\vec{J}\cdot\vec{E}-\nabla\cdot\vec{S}$ and $\partial u/ \partial t$ are both near zero. The balance of each of these terms matches qualitatively with both our simple theory and our PIC simulation results. However, since $\vec{J}\cdot\vec{E}$ is not balanced by $-\nabla\cdot\vec{S}$ elsewhere in the EDR, it is not clear if the balance of these two terms at/near the X-point is significant. {{(Note also that the exact agreement between $J_ME_M$ and $J_ME_{M,pred}$ requires a reconnection rate of 0.1, which is only a conical ``order of magnitude'' estimate rather than a known quantity.)}}

\section{Discussion}

We have investigated energy conversion in the central reconnection diffusion region by evaluating the source/loss ($\vec{J}\cdot\vec{E}$), flux divergence ($\nabla\cdot\vec{S}$), and time evolution ($\partial u/\partial t$) terms in Poynting's theorem. In theory, at the center of a symmetric steady-state laminar 2-d reconnecting current sheet, the energy conversion rate $J_ME_M$ balances the energy flux divergence $\partial (E_MB_L/\mu_0)/\partial N+\partial (E_MB_N/\mu_0)/\partial L$ such that $\partial u/\partial t=0$. We analyzed a 2.5-d particle-in-cell simulation of asymmetric reconnection and confirmed that, during a period where the reconnection rate was steady-state, energy balance ($\vec{J}\cdot\vec{E}=-\nabla\cdot\vec{S}\neq0$ such that $\partial u/\partial t=0$) is achieved in an area around the reconnection site in addition to the center of the reconnection site. For an MMS event, {{we found that the two sides of Poynting's theorem could be approximated uniquely and equivalently in such a way where the errors in each term were likely very small.}} Overall, we found that $\vec{J}\cdot\vec{E}$ and $\nabla\cdot\vec{S}$ did not balance one another {{as $-\nabla\cdot\vec{S}>\vec{J}\cdot\vec{E}>0$}}, leading to magnetic energy accumulation in the EDR. However, at the center of a reconnecting current sheet at/near the X-line, MMS observed energy balance similar to our basic theory for steady-state reconnection.

Our conclusion is that reconnection was not locally steady-state at the time when and place where it was observed by MMS. Given the strength of $J_NE_N$ and the co-located negative value of $\partial S_L/\partial L$, we suggest that this MMS-observed EDR may be better described by the picture of spatially oscillatory dissipation of \textit{Swisdak et al.} [submitted]. In their high-resolution 2.5 and 3-d PIC simulations, \textit{Swisdak} found that local fluctuations in the current sheet geometry can lead to the dissipation of the guide field $B_M$ component. If $B_M$ was being dissipated, then its strength would change with $L$ and lead to $\partial S_L/\partial L<0$, as was observed by MMS. This conclusion would be similar to that of \textit{Genestreti et al.} [submitted], which analyzed the same MMS event studied in this paper. \textit{Genestreti et al.} found that the form of the generalized Ohm's law was not consistent with 2-d laminar and steady-state reconnection.

Many open questions remain. Namely, can the observed energy imbalance can be replicated with 2-d laminar time-dependent reconnection? Otherwise, are these observations better described by 3-d and/or turbulent reconnection? These open questions should be addressed by future simulation and theory-driven studies. We also do not know what influence high-frequency waves have on the energy balance equation. There were no very-large-amplitude waves observed for this event, but wave generation, dampening, and propagation should almost certainly have at least some small role in governing the energy balance. Finally, for this MMS event, is it significant that $\vec{J}\cdot\vec{E}=-\nabla\cdot\vec{S}\neq0$ at the X-line but not on either side of the X-line, or is this a coincidence?

\acknowledgments
MMS data were obtained from the Science Data Center (SDC) at https://lasp.colorado.edu/mms/sdc/. The best-calibrated level 3 electric field data are available upon request. K.J. Genestreti was funded by the  FFG project number 847969. P. A. Cassak was funded by NASA Grant NNX16AG76G and NSF Grant AGS 1602769. Thanks to the entire MMS team that contributed to the success of the mission overall and the calibration of the instruments, without which this study would not have been possible. Thanks to Doctors N. Ahmadi, M. Akhavan-Tafti, P. Bourdin, L.-J. Chen, R.E. Ergun, S.A. Fuselier, R.B. Torbert, and M. Hesse for enlightening conversations and help with the MMS data. This event was identified during an International Space Science Institute (ISSI) meeting of the``MMS and Cluster observations of magnetic reconnection'' group. The study made use of the Space Physics Environment Data Analysis Software (SPEDAS) package.

\listofchanges

\end{document}